# Integrated near-field thermophotovoltaic device overcoming far-field blackbody limit

*Takuya Inoue,[1,*] Keisuke Ikeda,[2] Bongshik Song,[2,3] Taiju Suzuki,[2] Koya Ishino,[2] Takashi Asano,[2] and Susumu Noda[1,2*]*

[1] *Photonics and Electronics Science and Engineering Center, Kyoto University, Kyoto-daigaku-katsura, Nishikyo-ku, Kyoto 615-8510, Japan.*

[2] *Department of Electronic Science and Engineering, Kyoto University, Kyoto-daigaku-katsura, Nishikyo-ku, Kyoto 615-8510, Japan.*

[3] *Department of Electrical and Computer Engineering, Sungkyunkwan University, Suwon 16419, South Korea.*

*t_inoue@qoe.kuee.kyoto-u.ac.jp, snoda@kuee.kyoto-u.ac.jp

**Abstract**

Near-field thermal radiation transfer overcoming the far-field blackbody limit has attracted significant attention in recent years owing to its potential for drastically increasing the output power and conversion efficiency of thermophotovoltaic (TPV) power generation systems. Here, we experimentally demonstrate a one-chip near-field TPV device overcoming the far-field blackbody limit, which integrates a 20-μm-thick Si emitter and an InGaAs PV cell with a sub-wavelength gap (<140 nm). The device exhibits a photocurrent density of 1.49 A/cm$^2$ at 1192 K, which is 1.5 times larger than the far-field limit at the same temperature. In addition, we obtain an output power of 1.92 mW and a system efficiency of 0.7% for a 1-mm$^2$ device, both of which are one to two orders of magnitude greater than those of the previously reported near-field systems. Detailed comparisons between the simulations and experiments reveal the possibility of a system efficiency of >35% in the up-scaled device, thus demonstrating the potential of our integrated near-field TPV device for practical use in the future.

**Introduction**

Thermal radiation transfer between two objects that are separated by a sub-wavelength gap can be orders of magnitude larger than that in free space owing to the increase in the photonic density of states.[1-6] This concept has attracted significant attention in both fundamental science and various energy-related applications in recent years. The thermophotovoltaic (TPV) systems,[7-13] which convert heat into electricity by irradiating PV cells with thermal radiation from heated emitters, can significantly benefit from near-field thermal radiation transfer owing to the potential to increase output power density and conversion efficiency.[14-19] However, the benefit of near-field thermal radiation transfer has yet to be exploited in the real TPV systems because it is significantly challenging to realize a sub-wavelength gap and large temperature difference between a reasonably large emitter and PV cell while minimizing the thermal conduction loss. Although several proof-of-concept demonstrations of near-field TPV systems have been reported in recent years,[20-22] the generated electrical power in these systems was less than 1–10 µW owing to the small device size (<100 µm), and the system efficiency, which is defined by the ratio of the electrical output to the input heating power, was extremely low (<0.01%) owing to the enormous thermal conduction losses of the systems. In addition, these near-field systems involved external controllers such as piezoelectric actuators and NEMS actuators for the gap formation, which are not suitable for practical applications. To solve these issues, in our previous study,[23] we developed a near-field TPV device integrating a 2-µm-thick Si emitter with a side length of 500 µm and an InGaAs PV cell, and demonstrated the enhanced photocurrent compared to the corresponding far-field device. The thermal radiation intensity and the generated photocurrent in the above device, however, did not exceed the far-field blackbody limit at the same temperature, and the unintentional contact between the emitter and the PV cell increased the thermal conduction loss, thereby resulting in a relatively low output power (30 µW at 1065 K) and low system efficiency (0.05% at 1065 K).

In this paper, we demonstrate an integrated near-field TPV device that achieves the thermal radiation transfer overcoming the far-field blackbody limit and realize a drastic increase in both the output power and system efficiency. Our one-chip super-Planckian device is based on the realization of a sub-wavelength gap (<140 nm) between a high-temperature emitter (~1200 K) and a room-temperature PV cell without contact over a large area (1 mm$^2$), which is accomplished by designing a 20-µm-thick Si emitter with supporting beams that can relieve the thermal stress of the emitter while maintaining the mechanical robustness. In addition, an increase of the emitter thickness from 2 to 20 µm facilitates the realization of the near-field thermal radiation transfer beyond the blackbody limit owing to the increase in the photonic density of states. Using this device, we demonstrate a large photocurrent of 14.9 mA (density: 1.49 A/cm$^2$) at an emitter temperature of 1192 K, which is 1.5 times larger than the far-field blackbody limit at the same temperature. In addition, we obtain an output power of 1.92 mW (density: 0.192 W/cm$^2$) and a system conversion efficiency of 0.7%, both of which are greater than those of the previously reported near-field TPV systems[20-23] by one to two orders of magnitude. Furthermore, we theoretically predict that a system efficiency of >35% can be achieved in the up-scaled device with photon recycling, demonstrating the potential of our integrated near-field TPV device in practical applications.

**Results**

Figure 1(a) shows a bird's eye view and cross section of the proposed device: a 20-µm-thick Si thermal emitter with a side length of 1 mm was integrated on one side of an intermediate Si substrate while maintaining a sub-wavelength gap between them, and a thin InGaAs PV cell of the same size was integrated to the other side of the substrate. To suspend the millimeter-sized emitter while minimizing the tilt and the thermal conduction loss, we employed 10-μm-width supporting beams at the four corners instead of a single beam employed in our previous study.[23] Because the width of the supporting beams (10 μm) is smaller than the thickness (20 μm), these supporting beams can relieve the thermal stress of the emitter by in-plane deformation, which helps maintain a sub-wavelength gap between the emitter and the PV cell even at high temperatures (see Supporting Section 1 for details). In addition, compared to the thin-film Si thermal emitter ($t_e$=2 μm),[23] the 20-μm-thick Si thermal emitter enables thermal radiation transfer to overcome the far-field blackbody limit more easily owing to the increase of the photonic density of states inside the emitter. Figure 1(b) shows the calculated thermal radiation transfer spectra from the emitters with two different thicknesses (red line: $t_e$= 20 μm, blue line: $t_e$= 2 μm) to the InGaAs PV cell when the emitter temperature is 1200 K and the gap length is 150 nm. The details of the calculations are explained in Supporting Section 2. By increasing the thickness of the emitter, the radiation intensity below the bandgap wavelength of InGaAs ($\lambda_g$= 1.7 μm) is enhanced by a factor of 3, which exceeds the far-field blackbody limit (black line) in the entire near-infrared range at the same temperature. The thermal radiation transfer to the PV cell at longer wavelengths is relatively suppressed because the total thickness of the doped layers in the PV cell is only 2.5 μm, which is not sufficient for the doped carriers in the PV cell to induce strong free carrier absorption. It should be noted that the thermal radiation loss in the opposite side of the PV cell (not shown) increases with the emitter thickness but the loss can be reduced by placing the top reflector above the emitter for photon recycling.[19] Figure 1(c) shows the calculated photocurrent density of the PV cell as a function of the gap length; the dashed line shows the calculated photocurrent density for the far-field blackbody spectrum at the same temperature (hereafter, this value is referred to as the blackbody limit of the photocurrent). To obtain a photocurrent density that exceeds the blackbody limit, the gap length should be below 200 nm for the 20-μm-thick emitter, which ensures the tolerance to fabrication errors compared to that required for the 2-µm-thick emitter (below 100 nm).

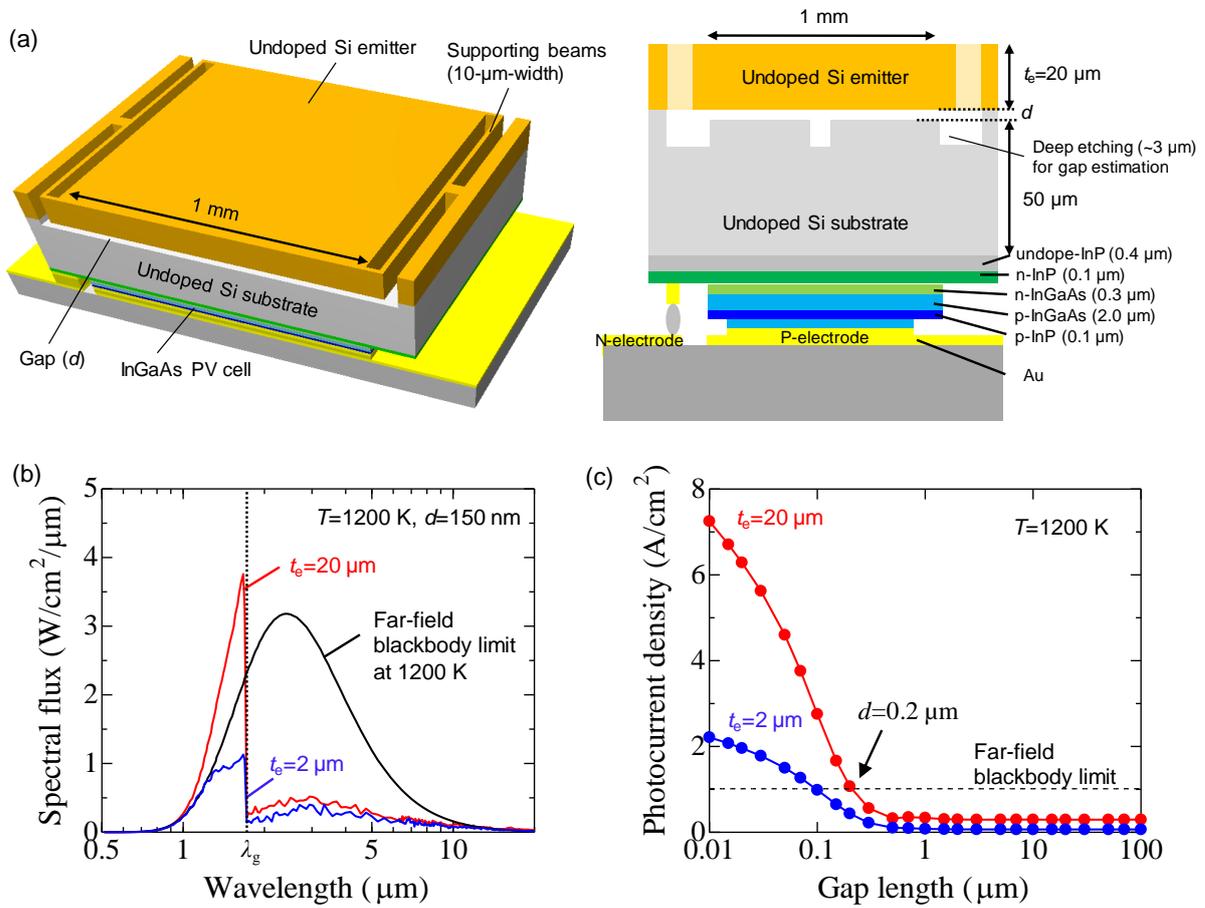

**Fig. 1. Schematics and calculated performances.** (a) Bird's eye view and cross section of near-field TPV device integrating 20-µm-thick Si thermal emitter and InGaAs PV cell. (b) Calculated thermal radiation transfer spectrum to the InGaAs PV cell for a 20-µm-thick (red) and 2-µm-thick (blue) Si thermal emitter when the gap length is 150 nm and the emitter temperature of 1200 K. Black line shows the far-field blackbody spectrum at 1200 K. (c) Calculated photocurrent density of the near-field TPV device with a 20-µm-thick (red) and 2-µm-thick (blue) Si thermal emitter at 1200 K. Dashed line shows the calculated photocurrent density for the far-field blackbody spectrum at the same temperature (blackbody limit).

Figure 2(a) shows a microscope image of the fabricated Si thermal emitter with a side length of 1 mm, which was integrated on top of the intermediate Si substrate via Si-Si bonding[24] (the details of the fabrication process are explained in Methods and Supporting Fig. S3). The gap length $d$ was controlled by the surface etching of the intermediate substrate before chip-to-chip bonding. Here, we fabricated one far-field TPV device (Device I, $d$~2900 nm) and two near-field TPV devices (Device II and III, $d$~150 nm), wherein the latter devices contain a small portion of 2900-nm-deep trenches with a side length of 0.1 mm for gap estimation. The microscope image of the fabricated 10-μm-width supporting beam is shown in the right panel. Because of the relatively strong mechanical strength of the 20-μm-thick Si slab, we can maintain the flatness of the emitter even with such elongated supporting beams. Figure 2(b) shows the microscope images of the fabricated InGaAs PV cells with a side length of 1 mm, which were integrated at the bottom of the intermediate substrate via plasma-assisted wafer bonding.[25] In this work, we investigated two types of PV cells; one (left panel, employed in Device I and II) has a comb-like p-type electrode (Au), while the other (right panel, employed in Device III) has a uniform p-type electrode. The comb-like electrode can reduce the absorption of sub-bandgap photons at the interface between the Au and the semiconductor, and thus, it is potentially suitable for high-efficiency TPV systems, but it leads to a higher series resistance of the PV cell owing to the low electrical conductance of the p-type semiconductor layer.

Before the near-field TPV experiment, we measured the in-plane distribution of the gap length of the fabricated three devices by varying the heating power of the Si emitter by laser irradiation (see Methods). We measured the reflection spectra at 5 × 5 points in each device by irradiating it with a broadband infrared light and estimated the gap length and the emitter temperature at each point from the Fabry-Perot interference of the device (the detailed procedure for the gap and temperature measurement is provided in Methods and Supporting

Section 4). Figure 2(c) shows the temperature dependence of the obtained gap lengths of the three devices, wherein the maximum, average, and minimum gap lengths within each device are shown in red, black, and blue, respectively. Although the obtained gap length was not uniform for each device [Fig. 2(c) and Supporting Fig. S4], the average gap length of each device remains almost the same for a wide range of emitter temperatures (300–1200 K). Among the three devices, Device II has the smallest average gap length (<140 nm at the emitter temperature of 1192 K), which satisfies the condition for exceeding the far-field blackbody limit shown in Fig. 1(c).

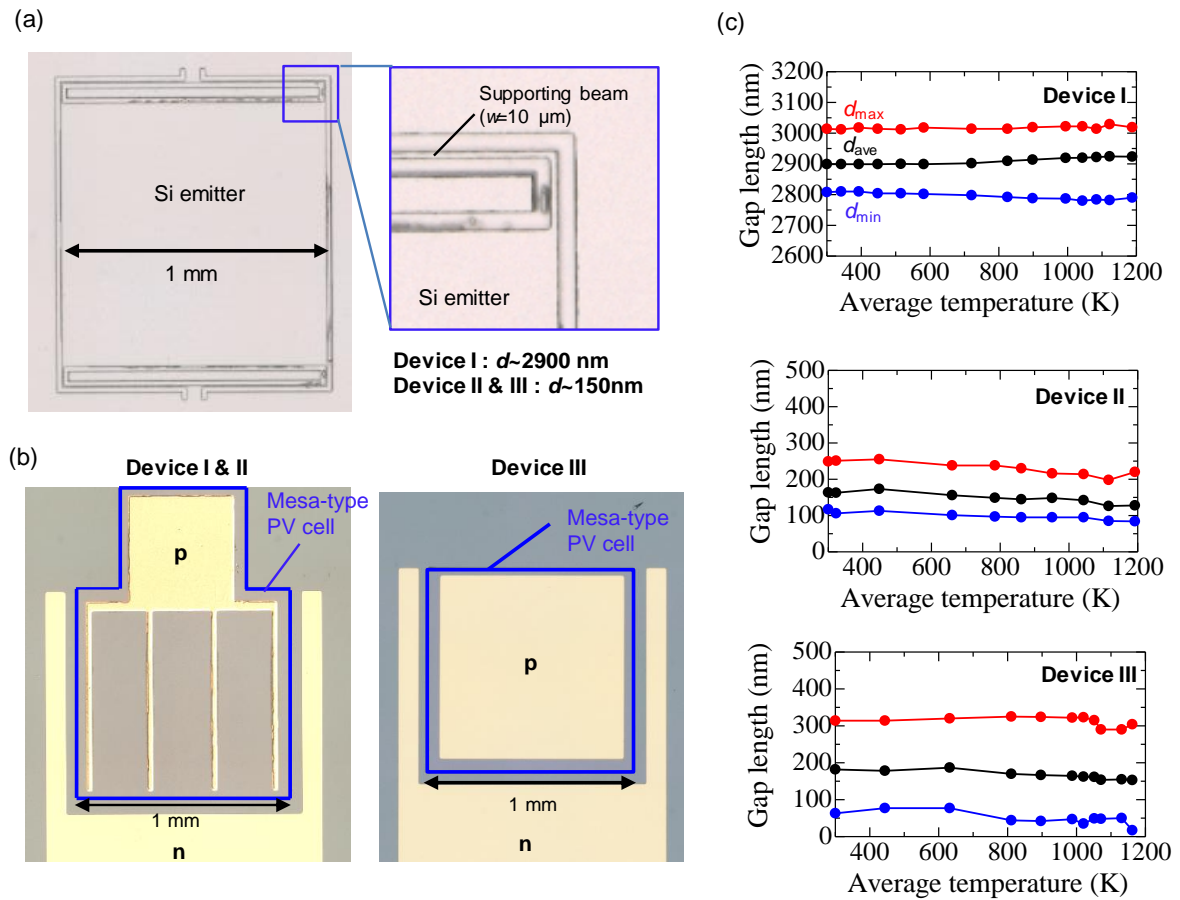

**Fig. 2. Fabricated devices.** (a) Microscope image of the fabricated Si emitter and enlarged view of the supporting beam. (b) Microscope image of the fabricated InGaAs PV cells with a comb-like p-type electrode (left) and a uniform p-type electrode (right). (c) Measured gap

lengths of the fabricated three near-field TPV devices as a function of the average temperature of the emitter.

Figures 3(a) and 3(b) show the measured current-voltage (I-V) characteristics of the far-field TPV device (Device I, $d_{ave}$~2900 nm) and the near-field TPV device (Device II, $d_{ave}$~140 nm) for various emitter temperatures. It should be noted that both devices contain an emitter and a PV cell with the same structures, and the only difference is the average gap length. Here, Device II yields much larger photocurrents than Device I. For example, at the same emitter temperature of 1043 K, Device II yields a short-circuit current (2.69 mA) that is 7.3 times larger than that of Device I (0.37 mA). It should be noted that in Device II, the short-circuit currents at 0 V at higher emitter temperatures (1116 and 1192 K) were less than the original photocurrents generated by the near-field thermal radiation transfer owing to the series resistance of the PV cell. Because the photocurrent densities in Device II are much higher than those of the typical solar cells, even a small series resistance induces a non-negligible internal forward bias on the p-n junction, and thus, a part of the generated photocurrent is internally consumed as a forward current (see Supporting Section 5). Therefore, the original photocurrent generated by the near-field thermal radiation transfer can be accurately measured by applying a sufficient reverse bias to cancel this internal forward bias. Figure 3(c) shows the measured and calculated photocurrent densities of the two devices at a reverse bias of −1 V as a function of the average emitter temperature. The measured photocurrent densities of the fabricated devices (red and blue triangles) agree well with the calculated ones (red and blue dashed lines). More importantly, the obtained photocurrent density in Device II exceeds the blackbody limit (black solid line) at an emitter temperature larger than 1050 K; for example, the obtained photocurrent density at 1192 K is 1.49 A/cm$^2$, which is 1.5 times larger than the blackbody limit at the same temperature (0.96 A/cm$^2$). This result is supported by the calculated near-field thermal radiation transfer spectrum shown in Fig. 3(d), wherein the spectral flux absorbed in the p-n junction exceeds the far-field blackbody limit (black line) in the entire near-infrared range below the bandgap wavelength of InGaAs ($\lambda_g$=1.7 μm).

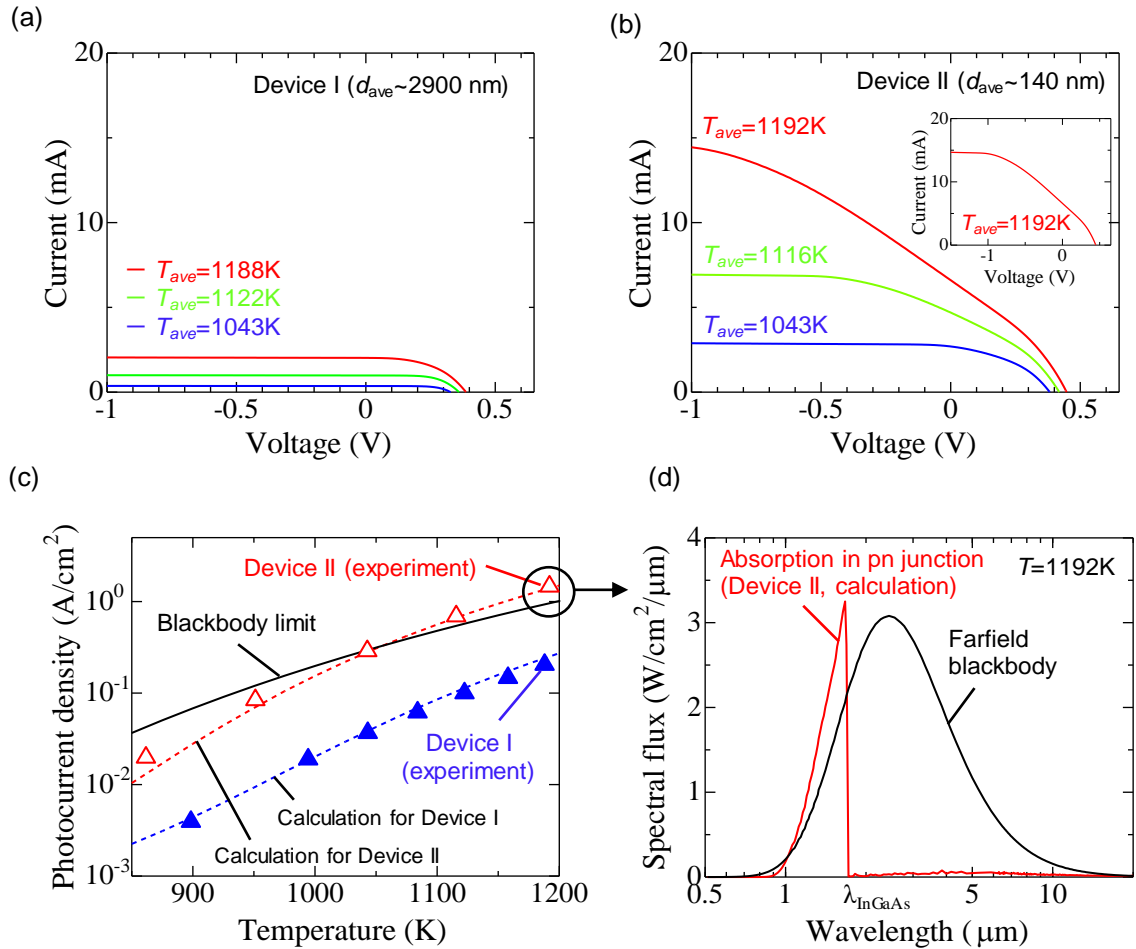

**Fig. 3. Photocurrent generation overcoming far-field blackbody limit.** (a) Measured current-voltage characteristics of the far-field TPV device (Device I) at three different emitter temperatures. (b) Measured current-voltage characteristics of the near-field TPV device (Device II) at three different emitter temperatures. As shown in the inset, the original photocurrent generated by the near-field thermal radiation transfer can be accurately measured by applying a sufficient reverse bias. (c) Measured (triangles) and calculated (dashed lines) photocurrent density of the two devices as a function of the average emitter temperature. Gap lengths of 3000 nm and 150 nm (except for deep trenches) were assumed in the calculations

for Device I and Device II, respectively. Black solid line shows the calculated blackbody limit for the InGaAs PV cell. (d) Calculated near-field thermal radiation transfer spectrum from the emitter to the p-n junction of the PV cell for Device II at 1192 K, wherein the gap length of 150 nm was assumed except for deep trenches.

Finally, we evaluated the actual system efficiency of our near-field TPV system by measuring both the input heating power of the emitter and the electrical output power. In this experiment, we characterized Device III, which has the lowest series resistance of the PV cell owing to the employment of the uniform p-type electrode [shown in the right panel of Fig. 2(b)]. The black dots in Fig. 4(a) show the relationship between the heating power of the emitter and the average temperature of the emitter in Device III. The black dashed line shows the calculated heat conduction loss through the supporting beams based on the temperature dependence of the thermal conductivity of Si.[26] When the temperature is lower than 650 K, the heating power of the emitter can be evaluated with the calculated heat conduction loss. The required heating power at higher temperatures exceeds the calculated conduction loss owing to the non-linear increase of the near-field thermal radiation power. The dashed lines (blue and red) show the calculation results obtained by taking the sum of the thermal conduction loss and the calculated total thermal radiation power from the emitter. Here, we varied the effective reflectance of the bottom electrode ($R_{\text{bottom}}$) to take into account the in-plane loss of thermal radiation in the finite-size device, where a portion of the waves reflected at the bottom electrode cannot return to the emitter (see Supporting Section 6). The experimental results agree well with the calculations with $R_{\text{bottom}} = 0.66$, which is lower than the reflectance of the ideal Au reflector ($R_{\text{Au}}=0.96$). Figure 4(b) shows the measured I-V characteristics of Device III (solid line) and Device II (dashed line) at almost the same heating power, wherein Device III yields a much larger electrical output power owing to the smaller series resistance of the PV cell.

Figure 4(c) shows the electrical output power density and system efficiency of Device III as a function of the emitter temperature. At an emitter temperature of 1162 K, we obtain the electrical power density of 0.192 W/cm$^2$. This corresponds to the absolute value of the electrical output power of 1.92 mW, which is two orders of magnitude larger than those of the previously demonstrated near-field TPV devices[20-23] owing to the larger device size (1 mm$^2$) and the higher emitter temperature. The maximum system efficiency of our device is 0.7 %, which is one order of magnitude higher than that of our previous near-field TPV device.[23] It should also be noted that the previous demonstrations of near-field TPV systems[20-22] resulted in lower system efficiencies (<0.01%) owing to the enormous thermal conduction loss.

    To increase the system efficiency, we should further decrease the gap length and increase the emitter temperature to enhance the near-field thermal radiation transfer below the bandgap wavelength of the PV cell. Figure 4(d) shows the calculated efficiency of the ideal near-field TPV devices with three gap lengths as a function of the emitter temperature. In this calculation, we assumed an infinite-size device with no conduction loss and ideal bottom reflectors ($R_{bottom}=R_{Au}=0.96$). We also modeled the dark current of the InGaAs PV cell by fitting the measured I-V characteristics (see Supporting Section 5) and neglected the series resistance for the ideal case. The solid lines show the efficiency of the device without a top Au reflector above the emitter as shown in Fig. 1(a), while the dashed lines show the efficiency of the device with a top Au reflector for photon recycling.[19] As seen in the figure, we can obtain the efficiency of ~20% without photon recycling by reducing the gap length to 75 nm at the temperatures higher than 1300 K. In addition, a higher conversion efficiency (>35%) can be achieved by placing the top Au reflector above the emitter and by recycling the thermal radiation opposite to the PV cell.

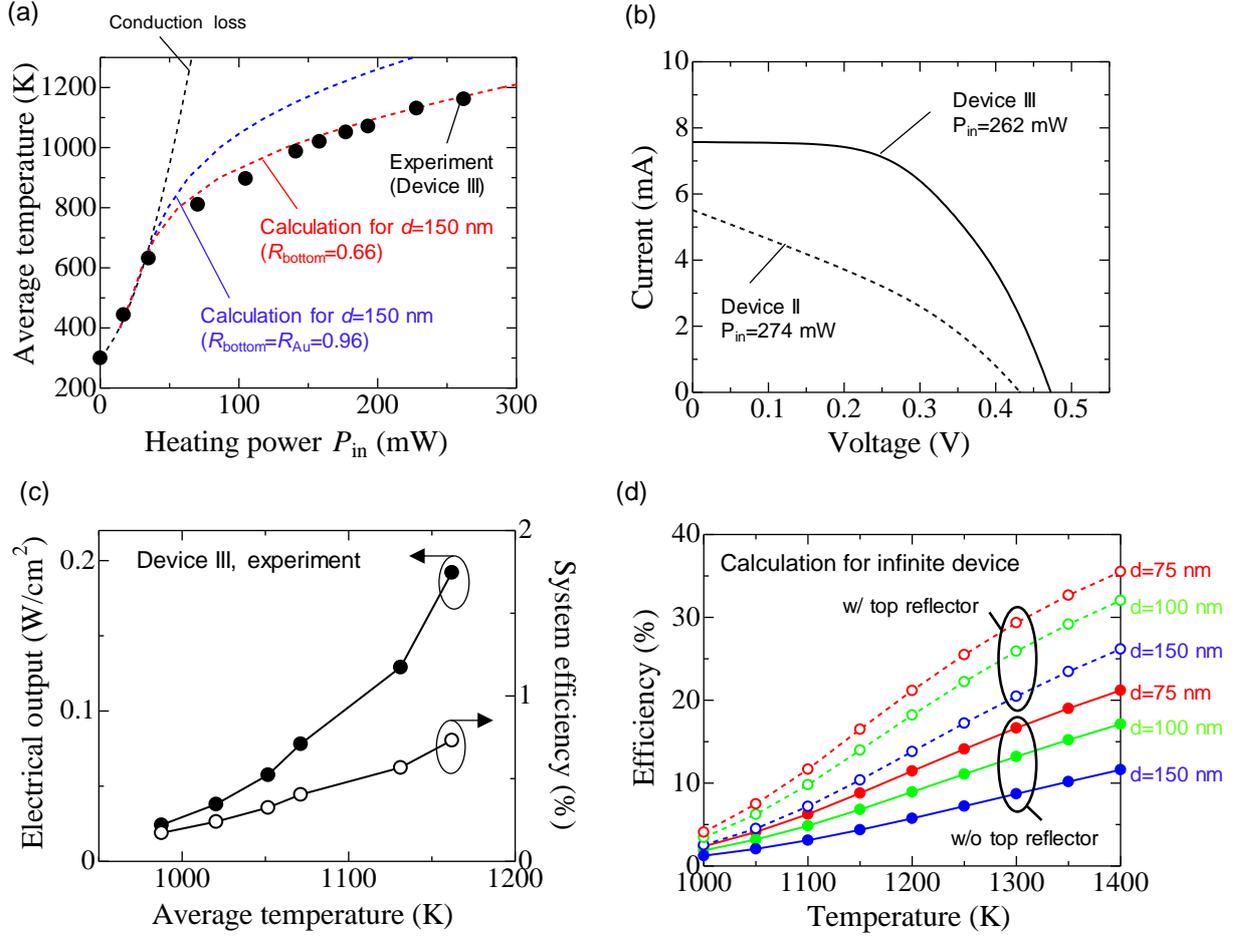

**Fig. 4. System efficiency of fabricated device and ideal device.** (a) Relationship between the heating power of the emitter ($P_{in}$) and the average temperature of the emitter in Device III. Black dashed line shows the calculated conduction loss of the emitter. Blue and red dashed lines show the calculation results obtained by taking the sum of thermal conduction loss and total thermal radiation power from the emitter, wherein we varied the effective reflectance of the bottom electrode ($R_{bottom}$). (b) Measured I-V characteristic of Device III at $P_{in}$=262 mW (solid line). I-V characteristic of Device II at almost the same $P_{in}$ is shown with a dashed line for comparison. (c) Measured output power density and system efficiency of Device III as a function of the emitter temperature. (d) Calculated system efficiency of an infinite-size near-field TPV device ($R_{bottom}=R_{Au}$, no conduction loss) with various gap lengths as a function of the emitter temperature. Solid and dashed lines show the efficiency of the system without and with a top Au reflector ($R_{top}=R_{Au}$) for photon recycling. In this calculation, we neglected the series resistance of the PV cell for ideal case.

**Discussion**

To experimentally realize the ideal system efficiencies shown in Fig. 4(d), several improvements in the device structure are necessary. First, we should further reduce the thermal conduction loss [black dashed line in Fig. 4(a)] and the additional thermal radiation loss due to the non-unity bottom reflectivity $R_{bottom}$ [shown with the power difference between the blue and red dashed lines in Fig. 4(a)]. Because these two losses stem from the finiteness of the device size, we can decrease them by scaling up the device size. Second, we should reduce the series resistance of the PV cell by optimizing the doping density of the n-doped contact layers to improve the fill factor of the I-V characteristics of the fabricated PV cell. Finally, to realize a smaller gap length at higher emitter temperatures, we should optimize the supporting structures of the emitter to further suppress the tilting and bowing of the emitter during heating.

In summary, we have developed a near-field TPV device integrating a 20-μm thick Si emitter and an InGaAs PV cell with a sub-wavelength gap (< 140 nm) without contact over a large area (1 mm$^2$). Using this device, we have demonstrated a large photocurrent (density) of 14.9 mA (1.49 A/cm$^2$) at 1192 K, which exceeds the far-field blackbody limit at the same temperature. In addition, by measuring both the input heating power and the electrical output power of our near-field TPV devices, we have obtained an output power of 1.92 mW and a system efficiency of 0.7%, which are greater than those of the previously reported near-field TPV systems by one to two orders of magnitude. We have also revealed that a high system efficiency (>35%) can be achieved in an up-scaled device by placing a top reflector above the emitter for photon recycling. Our one-chip super-Planckian devices will contribute to the full exploitation of near-field thermal radiation transfer in various applications including solar thermophotovoltaics and waste heat recovery.

**Methods**

**Sample preparation:** We prepared an SOI substrate with a 20-µm-thick top Si layer and a 1-µm-thick SiO$_2$ layer on a 650-µm-thick Si substrate (SOI-A) for the Si emitter, and another SOI substrate with a 50-µm-thick top Si layer and a 1-µm-thick SiO$_2$ layer on a 300-µm-thick Si substrate (SOI-B) for the intermediate Si substrate. The epitaxial wafer for InGaAs PV cells consisted of 400-nm undoped-InP/100-nm n-InP ($n_d$ = 2 × 10$^{18}$ cm$^{-3}$)/300-nm n-In$_{0.53}$Ga$_{0.47}$As ($n_d$ = 1 × 10$^{18}$ cm$^{-3}$)/2000-nm p-In$_{0.53}$Ga$_{0.47}$As ($n_a$ = 1 × 10$^{17}$ cm$^{-3}$)/100-nm p-InP ($n_a$ = 2 × 10$^{18}$ cm$^{-3}$)/300-nm p- In$_{0.53}$Ga$_{0.47}$As ($n_a$ = 2 × 10$^{18}$ cm$^{-3}$)/350-µm InP substrate. We first fabricated a Si emitter (1 × 1 mm) with four L-shaped supporting beams (width:10 µm, length:580 µm) on SOI-A by electron-beam (EB) lithography and cryogenic reactive ion etching (RIE). Next, we created a trench (Device I: 2900 nm, Device II and III: 150 nm) on SOI-B by RIE, in order to leave the gap between the emitter and the PV cell in the subsequent bonding process. For Devices II and III, we also created nine deeper trenches (~2900 nm) with a side length of 100 µm for gap estimation. After hydrophilizing the surfaces of the two substrates, we bonded them using a high-precision alignment and bonding system. The bonded sample was heated to 473 K in vacuum for 1 h and at 1273 K in Ar atmosphere for 1 h to increase the bonding strength. The upper Si substrate and SiO$_2$ layer were then removed by RIE and HF solution, respectively, to bare the 50-µm-thick intermediate Si substrate. The bonding of the intermediate Si substrate and the epi-structure for the PV cell was performed by oxygen plasma activation and 1-h post-annealing at 423 K in vacuum. The InP substrate was removed with HCl solution, and a mesa-type PV cell structure was formed by photolithography, metal deposition, and a lift-off process. The fabricated PV cell was then fixed on a supporting insulating substrate (Au/Ti/SiO$_2$/Si) by flip-chip bonding using a conductive adhesive (Ag paste). Finally, the Si substrate and $SiO_2$ layer above the emitter were removed by RIE and vapor HF etching, respectively, to bare the 20-µm-thick Si emitter. The schematic of each fabrication process is shown in Supporting Fig. S3.

**Near-field TPV experiment:** In the near-field TPV experiment, the Si emitters were heated by irradiation with a green laser ($\lambda$=532 nm) in a vacuum chamber (<1×10$^{-3}$ Pa). The heating power of the device, $P_{in}$, was calculated by considering the incident laser power and the theoretical reflectivity of Si at elevated temperatures. It should be noted that no laser light penetrates the emitter owing to sufficient absorption inside the 20-µm-thick emitter. To measure the emitter temperature and the gap length between the emitter and the intermediate substrate, we irradiated the device with broadband infrared light (wavelength range: 1000–1650 nm) and measured the reflection spectra for 5 × 5 points at 225-µm intervals in each device with a near-infrared spectrometer (Ocean Photonics, NIRQuest512). We then estimated the gap length and the emitter temperature at each point from the resonant wavelengths of the Fabry-Perot interference of the device (the details are explained in Supporting Section 4). The current-voltage characteristics of the PV cells were measured with a precision source/measure unit (Keysight, B2901A). The temperatures of the PV cells were kept at room temperature without any cooling system.

**Corresponding Author**

*E-mail: t_inoue@qoe.kuee.kyoto-u.ac.jp, snoda@kuee.kyoto-u.ac.jp

**Author Contributions**

TI, TA, and SN supervised the entire project. TI designed and fabricated samples with KI and BS. KI performed the experiments and analyzed the data with TI, ST, and KI. All the authors discussed the results and wrote the manuscript.

**Competing interests**

Authors declare that they have no competing interests.

**Acknowledgment**

This work was partially supported by a Grant-in-Aid for Scientific Research (17H06125) from the Japan Society for the Promotion of Science (JSPS).

**References**

1. D. Polder and M. V. Hove, Theory of radiative heat transfer between closely spaced bodies. *Phys. Rev. B* **4**, 3303–3314 (1971).

2. R. S. Ottens, V. Quetschke, S. Wise, A. A. Alemi, R. Lundock, G. Mueller, D. H. Reitze, D. B. Tanner, and B. F. Whiting, Near-Field Radiative Heat Transfer between Macroscopic Planar Surfaces. *Phys. Rev. Lett*. **107,** 014301 (2011).

3. B. Song, Y. Ganjeh, S. Sadat, D. Thompson, A. Fiorino, V. Fernandez-Hurtado, J. Feist, F. J. Garcia-Vidal, J. C. Cuevas, P. Reddy, and E. Meyhofer, Enhancement of near-field radiative heat transfer using polar dielectric thin films. *Nat. Nanotech*. **10**, 253–258 (2015).

4. R. St-Gelais, L. Zhu, S. Fan, and M. Lipson, Near-field radiative heat transfer between parallel structures in the deep subwavelength regime. *Nat. nanotech*. **11**, 515–520 (2016)

5. J. DeSutter, L. Tang, and M. Francoeur, A near-field radiative heat transfer device. *Nat. Nanotechnol*. **14**, 751–755 (2019).

6. H. Salihoglu, W. Nam, L. Traverso, M. Segovia, P. K. Venuthurumilli, and W. Liu, Near-field thermal radiation between two plates with sub-10 nm vacuum separation. *Nano Lett*. **20**, 6091–6096 (2020)

7. Yu Wei, Wenjuan Li, and Xianfan XuR. M. Swanson, A proposed thermophotovoltaic solar energy conversion system. *Proc. IEEE* **67**, 446–447 (1979).

8. M. Shimizu, A. Kohiyama, and H. Yugami, High-efficiency solar-thermophotovoltaic system equipped with a monolithic planar selective absorber/emitter. *J. Photonics Energy* **5**, 053099 (2015).

9. D. M. Bierman, A. Lenert, W. R. Chan, B. Bhatia, I. Celanović, M. Soljačić, and E. N. Wang, Enhanced photovoltaic energy conversion using thermally-based spectral shaping. *Nat. Energy* **1**, 16068 (2016).

10. T. Asano, M. Suemitsu, K. Hashimoto, M. De Zoysa, T. Shibahara, T. Tsutsumi, and S. Noda, Near-infrared–to–visible highly selective thermal emitters based on an intrinsic semiconductor. *Sci. Adv*. **2**, e1600499 (2016).

11. Z. Omair, G. Scranton, L. M. Pazos-Outón, T. P. Xiao, M. A. Steiner, V. Ganapati, P. F. Peterson, J. Holzrichter, H. Atwater, and E. Yablonovitch, Ultraefficient thermophotovoltaic power conversion by band-edge spectral filtering. *Proc. Natl. Acad. Sci. USA* **116**, 15356−15361 (2019).

12. D. Fan, T. Burger, S. McSherry, B. Lee, A. Lenert, and S. R. Forrest, Near-perfect photon utilization in an air-bridge thermophotovoltaic cell. *Nature* **586**, 237–241 (2020).

13. M. Suemitsu, T. Asano, T. Inoue, S. Noda, High-efficiency thermophotovoltaic system that employs an emitter based on a silicon rod-type photonic crystal. *ACS Photon*. **7**, 80−87 (2020).

14. M. Laroche, R. Carminati, and J.-J. Greffet, Near-field thermophotovoltaic energy conversion. *J. Appl. Phys*. **100**, 063704 (2006).

15. O. Ilic, M. Jablan, J. D. Joannopoulos, I. Celanovic, and M. Soljačić, Overcoming the black body limit in plasmonic and graphene near-field thermophotovoltaic systems. *Opt. Express* **20**, A367–A384 (2012).

16. J. K. Tong, W.-C. Hsu, Y. Huang, S. V. Boriskina, and G. Chen, Thin-film 'thermal well' emitters and absorbers for high-efficiency thermophotovoltaics. *Sci. Rep*. **5**, 10661 (2015).

17. M. P. Bernardi, O. Dupré, E. Blandre, P.-O. Chapuis, R. Vaillon, and M. Francoeur, Impacts of propagating, frustrated and surface modes on radiative, electrical and thermal losses in nanoscale-gap thermophotovoltaic power generators. *Sci. Rep*. **5**, 11626 (2015).

18. T. Inoue, K. Watanabe, T. Asano, and S. Noda, Near-field thermophotovoltaic energy conversion using an intermediate transparent substrate. *Opt. Express* **26**, A192–A208 (2018).

19. T. Inoue, T. Suzuki, K. Ikeda, T. Asano, and S. Noda, Near-field thermophotovoltaic devices with surrounding non-contact reflectors for efficient photon recycling. *Opt. Express* **29**, 11133–11143 (2021).

20. A. Fiorino, L. Zhu, D. Thompson, R. Mittapally, P. Reddy, and E. Meyhofer, Nanogap near-field thermophotovoltaics. *Nat. Nanotech*. **13**, 806–811 (2018).

21. G. R. Bhatt, B. Zhao, S. Roberts, I. Datta, A. Mohanty, T. Lin, J.-M. Hartmann, R. St-Gelais, S. Fan, and M. Lipson, Integrated near-field thermo-photovoltaics for heat recycling. *Nat. Commun.* **11**, 2545 (2020).

22. C. Lucchesi, D. Cakiroglu, J.-P. Perez, T. Taliercio, E. Tournié, P.-O. Chapuis, and R. Vaillon, Harnessing near-field thermal photons with efficient photovoltaic conversion. https://arXiv:1912.09394 (2019).

23. T. Inoue, T. Koyama, D. D. Kang, K. Ikeda, T. Asano, and S. Noda, One-chip near-field thermophotovoltaic device integrating a thin-film thermal emitter and photovoltaic cell. *Nano Lett.* **19**, 3948–3952 (2019).

24. S. Kawashima, M. Imada, K. Ishizaki, and S. Noda, High-precision alignment and bonding system for the fabrication of 3-D nanostructures. *J. Microelectromech. Syst.* **16**, 1140–1144 (2007).

25. D. Pasquariello and K. Hjort, Plasma-assisted InP-to-Si low temperature wafer bondings. *IEEE J. Sel. Top. Quant. Electron.* **8**, 118–131 (2002).

26. H. R. Shanks, P. D. Maycock, P. H. Sidles and G. C, Danielson, Thermal conductivity of silicon from 300 to 1400K. Phys. Rev. 130, 1743–1748 (1963).

# Supporting Information:
# Integrated near-field thermophotovoltaic device overcoming far-field blackbody limit

*Takuya Inoue,[1,*] Keisuke Ikeda,[2] Bongshik Song,[2,3] Taiju Suzuki,[2] Koya Ishino,[2] Takashi Asano,[2] and Susumu Noda[1,2*]*

[1] *Photonics and Electronics Science and Engineering Center, Kyoto University, Kyoto-daigaku-katsura, Nishikyo-ku, Kyoto 615-8510, Japan.*

[2] *Department of Electronic Science and Engineering, Kyoto University, Kyoto-daigaku-katsura, Nishikyo-ku, Kyoto 615-8510, Japan.*

[3] *Department of Electrical and Computer Engineering, Sungkyunkwan University, Suwon 16419, South Korea.*

*t_inoue@qoe.kuee.kyoto-u.ac.jp, snoda@kuee.kyoto-u.ac.jp

## 1. Simulation of temperature distribution and vertical displacement of emitter

As described in the main text, the 20-μm-thick emitter in our near-field TPV device is suspended with 10-μm-width beams at the four corners. In this structure, the narrow supporting beams can relieve the thermal stress of the emitter at high temperatures by an in-plane deformation. To verify the feasibility of the aforementioned factor, we calculated the temperature distribution and vertical displacement distribution of the heated emitter by the finite element method (COMSOL Multiphysics). To simulate laser heating in the experiment, we fixed the temperature of the intermediate substrate to 300 K and provided a circular heat flux ($D = 0.5$ mm) on the top surface of the 1-mm emitter. In the simulation, we considered the temperature dependence of the total thermal radiation power from the emitter, which was calculated beforehand using the fluctuation-dissipation theorem (see Supporting Section 2) by assuming a gap length of 100 nm. Figure S1(a) shows the calculated temperature distribution at the average emitter temperature of 1300 K. The in-plane variation of the temperature is within ±40 K except in the vicinity of the supporting beam. Figure S1(b) shows the calculated vertical displacement at the same temperature. Although the entire region of the emitter goes down by ~30 nm due to its own weight, the in-plane variation of the vertical displacement is negligible, which validates the effectiveness of our design.

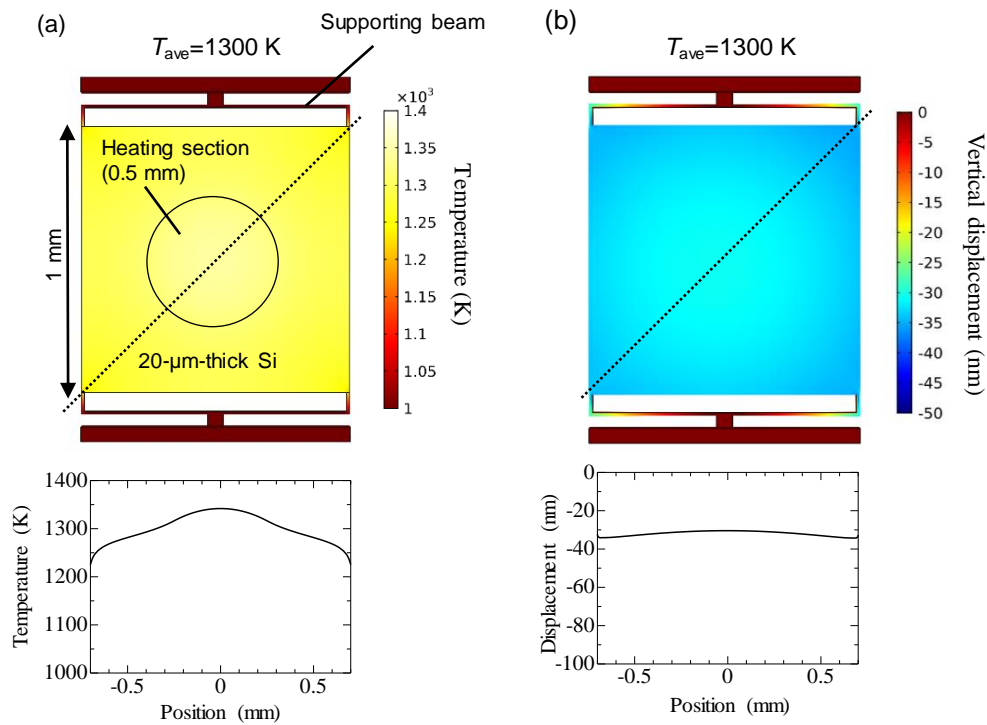

**Fig. S1. Calculated temperature and vertical displacement distributions.** (a) Temperature distribution of the 20-μm-thick emitter at the average temperature of 1300 K. (b) Vertical displacement distribution of the 20-μm-thick emitter at the average temperature of 1300 K.

## 2. Simulation model of near-field thermal radiation transfer

The simulation model of our near-field TPV device is shown in Fig. S2(a). In our simulations, the optical constants of undoped Si at high temperatures were taken from Refs. S1–S3, wherein the changes in bandgap energy and intrinsic carrier concentrations at elevated temperatures were considered. Figure S2(b) shows the theoretical absorption coefficient spectra of undoped Si at various temperatures. The absorption coefficient of Si in the near-infrared range increases as the temperature rises owing to the redshift of the bandgap wavelength, and thus, the near-infrared thermal radiation transfer exceeds the blackbody limit as the temperature increases. The sub-bandgap absorption coefficient caused by the thermally excited intrinsic carriers also increases as the temperature rises. The optical constants of InGaAs and InP at the near-infrared range were taken from Refs. S4 and S5, and those at the far-infrared range were modeled by Lorentz-Drude dispersions.[S6] The dielectric function of Au was described by Drude's model, wherein the plasma frequency and damping frequency were taken from Ref. S7. In the calculation of Fig. 4(a) in the main manuscript, we intentionally increased the damping frequency of Au to simulate the lower effective reflectance for the bottom electrode.

The calculation of the thermal radiation spectra transferred from the Si thermal emitter to each layer of the PV cell was performed using a method that combines the fluctuation–dissipation theorem and the transfer matrix method.[S8] The thermal radiation transfer spectrum between the emitter and each layer of the PV cell is given by

$$S(\lambda) = \frac{1}{4\pi^2} \frac{2\pi c}{\lambda^2} \left[ \Theta(\lambda, T_{emitter}) - \Theta(\lambda, T_{PV}) \right] \times \int_0^\infty \left[ Z_{TE}(\lambda, k) + Z_{TM}(\lambda, k) \right] k dk$$
$$\Theta(\lambda, T) = \frac{hc/\lambda}{\exp(hc/\lambda kT) - 1}$$
(S1)

where $Z_{TE}(\lambda, k) \left[ Z_{TM}(\lambda, k) \right]$ is defined as the exchange function between the emitter and each layer of the PV cell for the TE-polarized (TM-polarized) plane wave with an in-plane wavenumber $k$ and a wavelength $\lambda$, which can be calculated by using transfer matrix method. The spectral flux of the interband absorption $P_{\text{interband}}(\lambda)$, which contributes to the generation of electron-hole pairs, was calculated by taking the sum of $S(\lambda)$ for the p-InGaAs and n-InGaAs layers of the PV cell below the bandgap wavelength of the PV cell. The photocurrent density in the PV cell was calculated as follows:

$$J_p = e \int_{\lambda < \lambda_g} \frac{P_{\text{interband}}(\lambda)}{hc/\lambda} d\lambda. \tag{S2}$$

Here, we assumed that each photon absorbed by the interband transition in the InGaAs pn junction generates an electron-hole pair, which is applicable for our InGaAs PV cells. The total near-field thermal radiation power from the emitter was calculated by integrating $S(\lambda)$ in the entire wavelength range and taking the sum for all the layers, wherein we also considered the absorption in the bottom Au electrode and the far-field thermal radiation in the opposite direction.

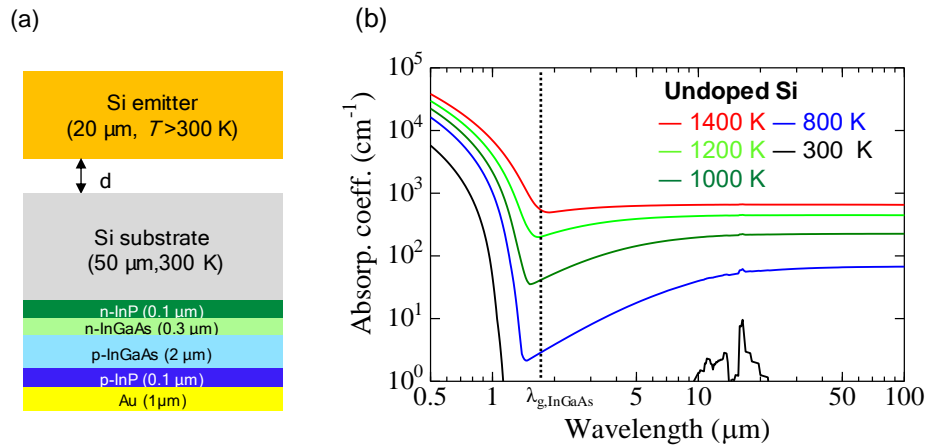

**Fig. S2. Simulation model.** (a) Schematic of the proposed one-chip near-field TPV device. (b) Absorption coefficient of undoped Si as a function of temperature.

## 3. Fabrication of near-field TPV device

Figure S3 shows the fabrication process of our near-field TPV device. First, a Si emitter (1 mm × 1 mm) with four L-shaped supporting beams (width:10 μm, length:580 μm) were formed on a 20-μm-thick Si layer of a silicon-on-insulator (SOI) substrate by electron-beam (EB) lithography and cryogenic reactive ion etching (RIE) [left panels of Fig. S3(a)]. Next, we created a trench (Device I: 2900 nm, Device II and III: 150 nm) on a 50-μm-thick Si layer of another SOI substrate (intermediate substrate) by RIE [right panels of Fig. S3(a)], in order to leave a gap between the emitter and the PV cell in the subsequent bonding process. For Devices II and III, we also created nine deeper trenches (~2900 nm) with a side length of 100 μm for gap estimation. After hydrophilizing the surfaces of the two substrates, we bonded them using a high-precision alignment and bonding system [Fig. S3(b)]. The bonded sample was heated to 473 K in vacuum for 1 h and at 1273 K in Ar atmosphere for 1 h to increase the bonding strength. The upper Si substrate and $SiO_2$ layer were then removed with RIE and HF solution, respectively, to bare the 50-μm-thick intermediate Si substrate [Fig. S3(c)]. The bonding of the intermediate Si substrate and the epi-structure for the PV cell (n-InP/n-$In_{0.53}Ga_{0.47}As$/p-$In_{0.53}Ga_{0.47}As$/p-InP/p-$In_{0.53}Ga_{0.47}As$) was performed by oxygen plasma activation and 1-h post-annealing at 423 K in vacuum [Fig. S3(d)]. The InP substrate was removed with HCl solution [Fig. S3(e)], and a mesa-type PV cell structure was formed by photolithography, metal deposition, and a lift-off process [Fig. S3(f)]. The fabricated PV cell was then fixed on a supporting insulating substrate (Au/Ti/$SiO_2$/Si) by flip-chip bonding using a conductive adhesive (Ag paste) [Fig. S3(g)]. Finally, the Si substrate and $SiO_2$ layer above the emitter were removed by RIE and vapor HF etching, respectively, to bare the 20-μm-thick Si emitter [Fig. S3(h)].

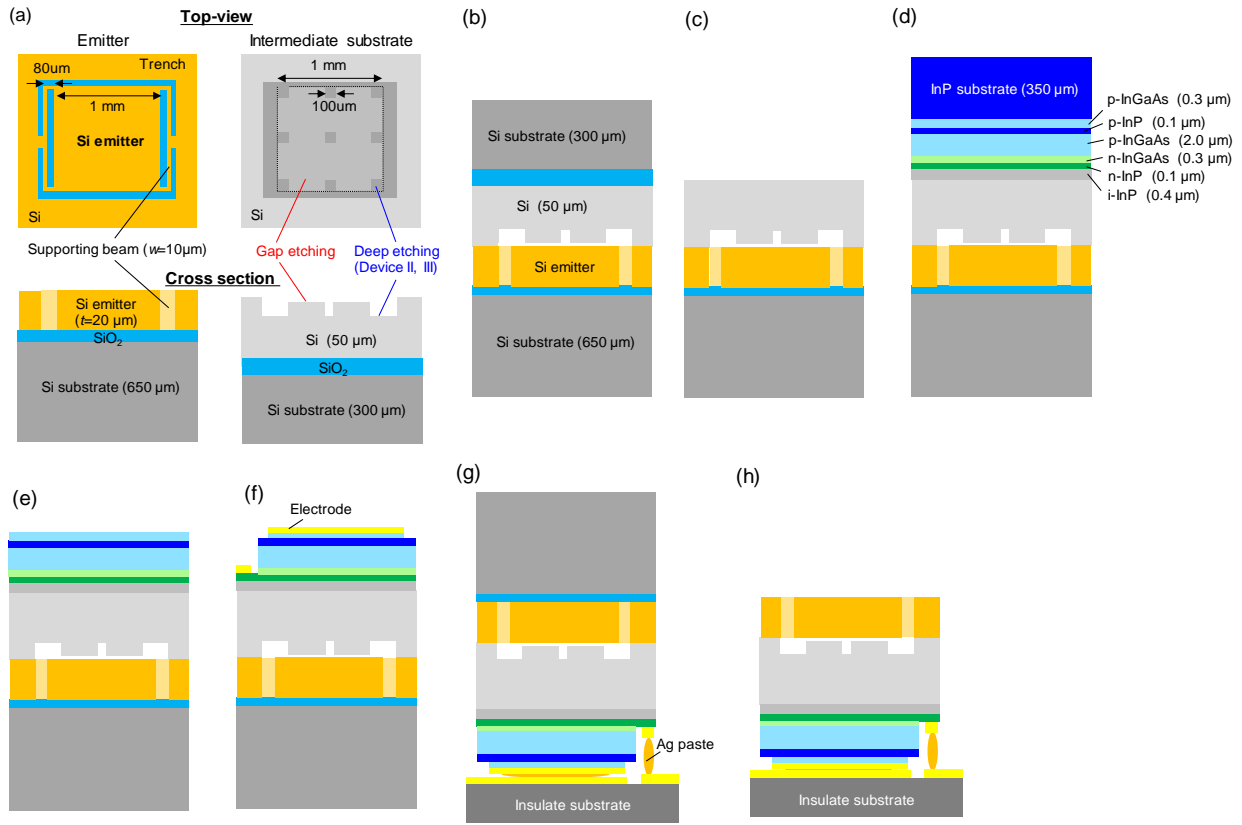

**Fig. S3. Fabrication process of near-field TPV device.** (a) Patterning of 20-μm-thick emitter and intermediate substrate by EB lithography and RIE. (b) Bonding of the emitter and the intermediate substrate. (c) Removal of the Si substrate and SiO$_2$ layer to bare the intermediate substrate. (d) Bonding of the intermediate substrate and the epi-structure for the PV cell. (e) Removal of the InP substrate. (f) Fabrication of a mesa-type PV cell. (g) Flip-chip bonding on an insulate substrate. (h) Removal of the Si substrate and SiO$_2$ layer to bare the emitter.

## 4. Simulation of temperature distribution and vertical displacement of emitter

As briefly explained in the main text, we estimated the emitter temperature and actual gap distance from the measurement of the reflection spectra for broadband infrared light ($\lambda$ =1000–1650 nm). Figure S4(a) shows the schematic of the reflection measurement, wherein we measured the reflection spectra at 5 × 5 points in each device. Figure S4(b) shows the example of the measured spectra at various heating powers for Device II, where the Fabry-Perot interferences in the 20-μm-thick Si (indicated with 'A') and that in the vacuum gap (indicated with 'B') clearly appeared. The former interferences redshift as the temperature rises due to the increase in the refractive index of Si, while the latter does not change because its resonant wavelength is determined by the vacuum gap length $d$ as follows;

$$\lambda_m = \frac{2d}{m}, \tag{S3}$$

where $m$ is the order of the Fabry-Perot interference [$m$ = 4 in the case of Fig. S4(b)]. Therefore, the gap length $d$ at each point is easily determined by using Eq. S3. It should be noted that the Fabry-Perot interferences at the gap $d$ < 500 nm do not appear in the measured wavelength range. Therefore, for Devices II and III, we first estimated the gap in the deeply etched sections ($d$~2900 nm) and then calculated the gap in the other sections by linear interpolation considering the difference of the etching depth (2760 nm ± 10 nm). Thus, the uncertainty of the estimated gap at each point was within ± 10 nm.

To determine the temperature of the emitter, we compared the measured reflection spectra with the calculated ones. In this calculation, we considered the wavelength and temperature dependence of the refractive index of Si in the near-infrared range.[S9,S10] Figure S4(c) shows the calculated reflectivity spectra assuming different emitter temperatures. The resonant wavelengths of the Fabry-Perot interferences show excellent agreement with the measured ones in Fig. S4(b). From this comparison, we determined the emitter temperature at each point with an accuracy of ± 10 K. Figures S4(d)–S4(f) show the obtained temperature and gap distributions of the fabricated devices at the maximum heating power. The emitter temperature was highest in the center and lowest in the vicinity of the supporting beams. The maximum temperature difference within the emitter was ~100 K, which was comparable to the calculated results [Fig. S1(a)]. The obtained gap distributions indicate that the emitters of these three devices show a small degree of tilting or bowing probably due to the residual stress during the fabrication process. Such tilting/bowing should be suppressed to realize a smaller gap length between the emitter and the PV cell in the future.

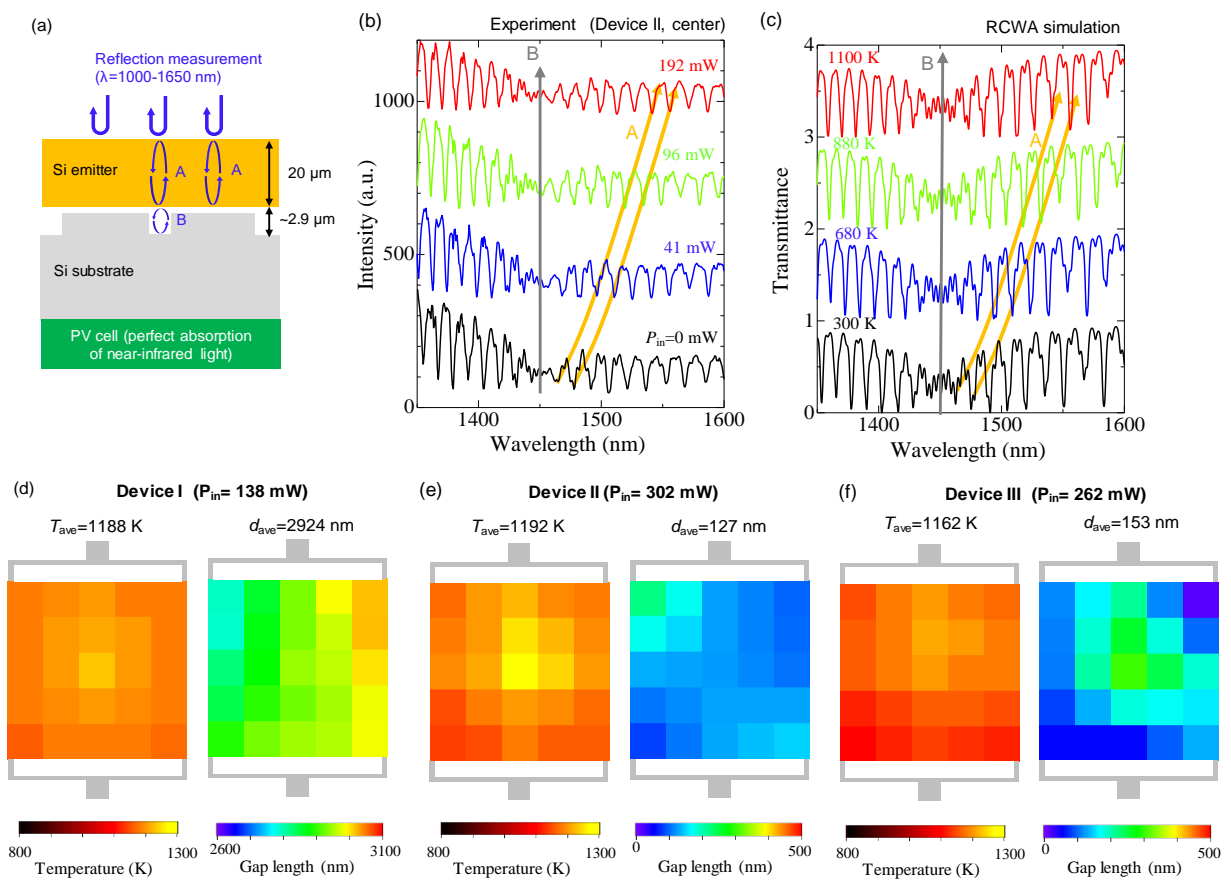

**Fig. S4. Measurement of temperature and gap distributions.** (a) Method for estimating the temperature of the thermal emitter and the gap between the emitter and the intermediate substrate. (b) Example of the measured reflection spectra of the Si thermal emitter (Device II) with various heating powers. (c) Calculated reflectivity spectra assuming different emitter temperatures. (d)(e)(f) Measured temperature (left) and gap (right) distributions of Device I, II,

III at the maximum heating power. Depth difference for the deeply etched sections of Device II and III (2760 nm) is subtracted in (e) and (f).

## 5. Theoretical modeling of PV cell

The current density in the PV cell is given by the following equation:

$$J(V) = J_p - J_{dark}(V) \quad , \tag{S4}$$

where $J_p$ is the photocurrent density calculated by Eq. (S2) in Supporting Section 2, and $J_{dark}(V)$ is the dark current density of the PV cell at an applied voltage $V$. For the calculation of $J_{dark}(V)$, we used the following approximation;

$$J_{dark}(V) = J_0 \left[ \exp\left(\frac{eV_{int}}{nkT_{cell}}\right) - 1 \right] + \frac{V_{int}}{R_{sh}S_{cell}}, \tag{S5}$$

where $J_0$ is the reverse saturation current density, $V_{int}$ is the internal voltage at the p-n junction, $n$ is the ideality factor, $k$ is the Boltzmann constant, $T_{cell}$ is the temperature of the PV cell, $R_{sh}$ is the shunt resistance [relatively large (~MΩ) for the fabricated PV cell], and $S_{cell}$ is the area of the PV cell. The $V_{int}$ is larger than the applied voltage $V$ even with a relatively small series resistance because the photocurrent density in the near-field TPV cell is much higher than that in the conventional solar cell. In the case of the millimeter-sized TPV cell, $V_{int}$ is not uniform inside the device due to the in-plane resistance of the contact layer. Figures S5(a) and S5(b) show the schematic and the one-dimensional equivalent circuit model of the fabricated TPV cells with comb-like and uniform bottom electrodes, respectively. Here, $j_p$ and $j_i$ ($i=1\sim N$) denote the photocurrent density and the dark current density at each position, $R_s$ denotes the series resistance outside the mesa structure, and $r_n$ and $r_p$ denote the in-plane resistance of the n-type and p-type contact layer per unit length, respectively. Because the mobility of the holes in p-type semiconductors is lower than that of the electrons in n-type semiconductors by one to two orders of magnitude, $r_p$ is greater than $r_n$, which results in a lower fill factor of the PV cell with the comb-like electrodes. By using these equivalent circuit models, we performed the fitting of the measured I-V characteristics of the fabricated devices and extracted the fitting parameters ($J_0$, $n$, $R_s$, $r_n$, $r_p$). The results of the fitting for three I-V curves with different photocurrents for Devices II and III are shown in Fig. S5(c) and S5(d), respectively. For each device, we successfully reproduced all the three I-V curves by using the same fitting parameters shown in each figure.

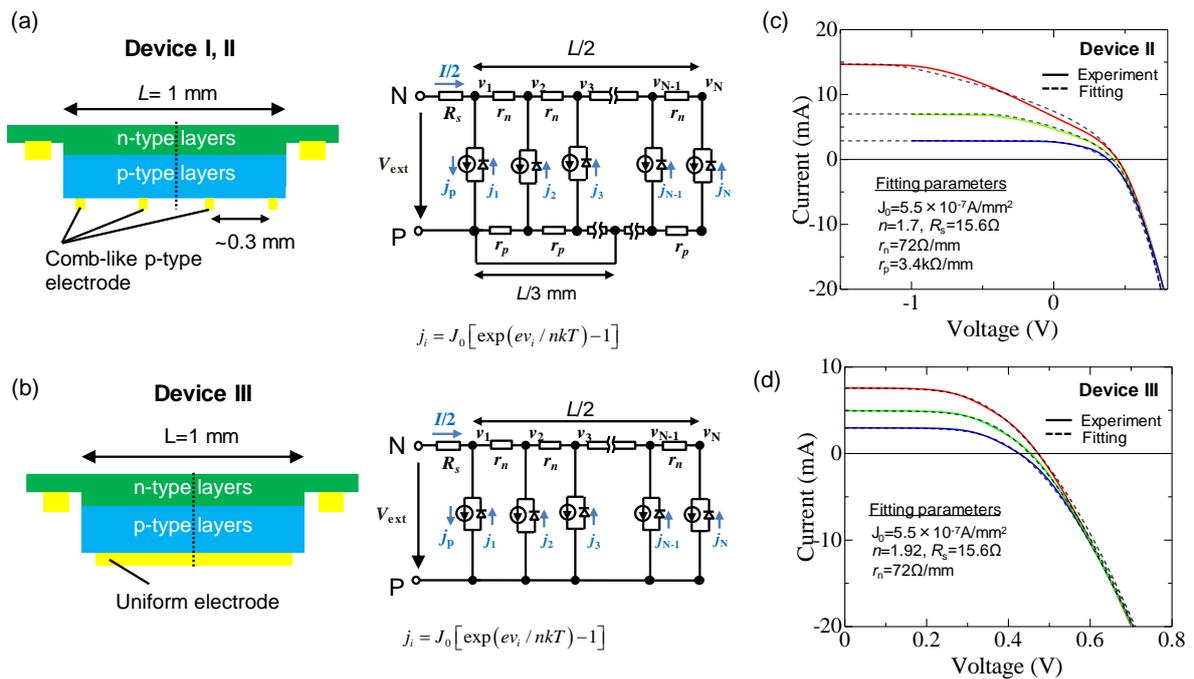

**Fig. S5. Modeling of the fabricated PV cells.** (a) Schematic and equivalent 1D circuit model of the fabricated PV cell with the comb-like p-type electrode. (b) Schematic and equivalent 1D circuit model of the fabricated PV cell with the uniform p-type electrode. (c)(d) Comparison between the measured and fitted I-V characteristics for Device II and III. Fitting parameters are shown in the figure.

## 6. Effective bottom reflectivity in finite-size device

In the simulation for near-field thermal radiation transfer (Supporting Section 2), we assume that the in-plane size of the system is infinite so that all of the thermal radiation reflected at the bottom of the PV cell returns to the emitter. It should be noted, however, that such a model is valid only when the in-plane size of the device $L$ [shown in Fig. S6(a)] is sufficient as compared to the vertical distance between the emitter and the bottom electrode ($d_{total}$~50 μm). The ratio of near-field thermal radiation which is reflected back to the finite-size emitter can be approximated from the simple calculation of the view factor of the emitter, as shown in Fig. S6(b). Figure S6(c) shows the calculated view factor of the emitter as a function of the in-plane size of the device $L$. The calculated view factor for a 1-mm device is ~0.8, which indicates that the effective reflectance of the bottom electrode $R_{bottom}$ is less than 0.8 (note that $R_{bottom}$ also depends on the absorptivity of the Au electrode in the fabricated device). In the calculation of the total thermal radiation power of the fabricated device [red dashed line in Fig. 4(a)], we replaced the bottom Au electrode with a virtual less-reflective material with an intentionally increased damping frequency to simulate the reduced bottom reflectivity, and found that the calculation with $R_{bottom}$=0.66 fitted the experimental results well. From the result shown in Fig. S6(c), the in-plane thermal radiation loss can be further suppressed by enlarging the emitter size to > 2 mm, or by reducing the thickness of the intermediate substrate to < 25 μm.

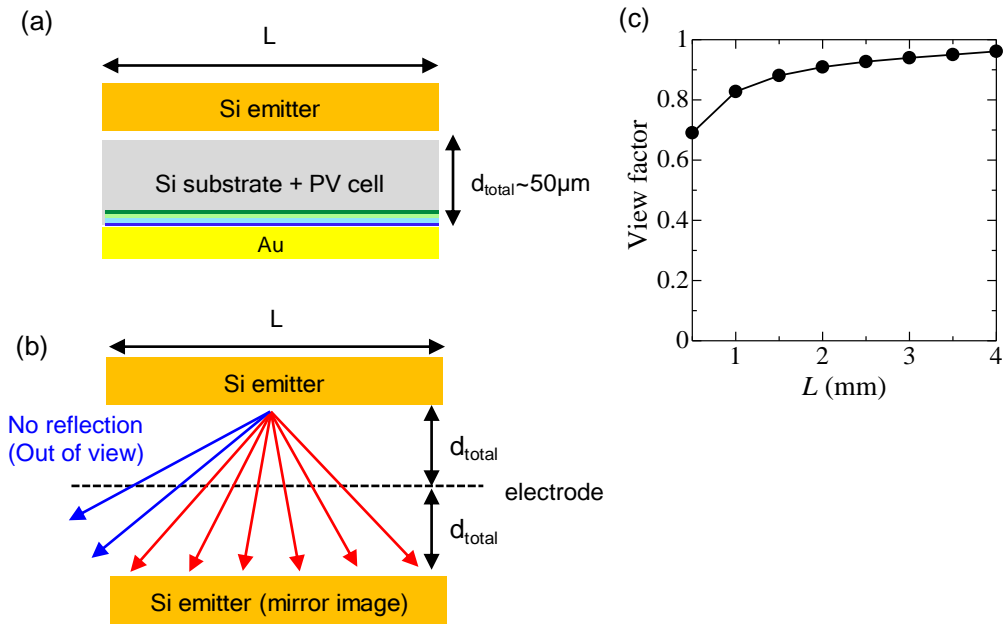

**Fig. S6. Effective reflectance of the bottom electrode.** (a) Schematic of a near-field TPV system with a bottom electrode. (b) Simulation model of a view factor of the emitter. (c) Calculated view factor as a function of in-plane size of the TPV device.

**References**

S1. G. E. Jellison, Jr. and D. H. Lowndes, Optical absorption coefficient of silicon at 1.152 µ at elevated temperatures. *Appl. Phys. Lett.* **41**, 594–596 (1982).

S2. P. J. Timans, Emissivity of silicon at elevated temperatures. *J. Appl. Phys.* **74**, 6353–6364 (1993).

S3. C. J. Fu and Z. M. Zhang, Nanoscale-radiation heat transfer for silicon at different doping levels. *Int. J. Heat Mass Transfer* **49**, 1703–1718 (2006).

S4. M. Munoz, T. M. Holden, F. H. Pollak, M. Kahn, D. Ritter, L. Kronik, and G. M. Cohen, Optical constants of $In_{0.53}Ga_{0.47}As/InP$: Experiment and Modeling. *J. Appl. Phys.* **92**, 5878–5885 (2002).

S5. S. Adachi, Optical dispersion relations for GaP, GaAs, GaSb, InP, InAs, InSb, AlxGa1−xAs, and In$_{1-x}$Ga$_x$As$_y$P$_{1-y}$. *J. Appl. Phys.* **66**, 6030–6040 (1989).

S6. D. J. Lockwood, G. Yu, and N. L. Rowell, Optical phonon frequencies and damping in AlAs, GaP, GaAs, InP, InAs and InSb studied by oblique incidence infrared spectroscopy. *Solid State Commun*. **136**, 404–409 (2005).

S7. M. A. Ordal, L. L. Long, R. J. Bell, S. E. Bell, R. R. Bell, R. W. Alexander, and C. A. Ward, Optical properties of Al, Co, Cu, Au, Fe, Pb, Ni, Pd, Pt, Ag, Ti, and W in the infrared and far infrared. *Appl. Opt*. **22**, 1099–1119 (1983).

S8. T. Inoue, T. Asano, and S. Noda, Near-field thermal radiation transfer between semiconductors based on thickness control and introduction of photonic crystals. *Phys. Rev*. B **95**, 125307 (2017).

S9. H. H. Li, Refractive index of silicon and germanium and its wavelength and temperature derivatives. *J. Phys. Chem. Ref. Data* **9**, 561–658 (1980).

S10. J. A. McCaulley, V. M. Donnelly, M. Vernon, and I. Taha, Temperature dependence of the near-infrared refractive index of silicon, gallium arsenide, and indium phosphide. *Phys. Rev. B* **49**, 7408–7417 (1994).